\documentclass[preprint,aps,amsmath,nofootinbib,tightenlines]{revtex4}
\usepackage{graphicx}
\usepackage{bm}
\usepackage{epsfig}
\usepackage{graphics}

\newif\ifpdf
\ifx\pdfoutput\undefined
\pdffalse % we are not running PDFLaTeX
\else
\pdfoutput=1 % we are running PDFLaTeX
\pdftrue
\fi

\def\Dsl{\hbox{/\kern-.6000em D}} %roman D

\def\dsl{\,\raise.15ex\hbox{/}\mkern-13.5mu D}

\def\psip#1{\psi_{\mathbf{#1}}}

\def\ltap{\ \raise.3ex\hbox{$<$\kern-.75em\lower1ex\hbox{$\sim$}}\ }
\def\gtap{\ \raise.3ex\hbox{$>$\kern-.75em\lower1ex\hbox{$\sim$}}\ }
\def\OMIT#1{}

\def\lsim{\mathrel{\raise.3ex\hbox{$<$\kern-.75em\lower1ex\hbox{$\sim$}}}}
\def\gsim{\mathrel{\raise.3ex\hbox{$>$\kern-.75em\lower1ex\hbox{$\sim$}}}}

\newcommand{\nn}{\nonumber}
\newcommand{\ppp}{\mbox{$({\mathbf p'}-{\mathbf p})^2$}}

\newcommand{\two}{{\rm 2}}

\newcommand{\bmk}{\mathbf k}
\newcommand{\bmp}{\mathbf p}

\newcommand{\bmq}{\mathbf q}

\newcommand{\bmA}{\mathbf A}

\newcommand{\bmE}{\mathbf E}

\newcommand{\bmD}{\mathbf D}

\catcode`\@=11
\def\slash{\mathpalette\make@slash}
\def\make@slash#1#2{\setbox\z@\hbox{$#1#2$}%
  \hbox to 0pt{\hss$#1/$\hss\kern-\wd0}\box0}
\catcode`\@=12 % @ signs are no longer letters
\begin{document}
%%%%%%%%%%%%%%%%%%%%%%%%%%%%%%%%%%%%%%%%%%
%Some more stuff to get graphics to work
\ifpdf
\DeclareGraphicsExtensions{.pdf, .jpg}
\else
\DeclareGraphicsExtensions{.eps, .jpg}
\fi
%%%%%%%%%%%%%%%%%%%%%%%%%%%%%%%%%%%%%%%%%%

%%%%%%%%%%%%%%%%%%%%%%%%%%%%%%%%%%%%%%%%%%
%Define Title, Author, Address, Preprint#

\preprint{ \vbox{ \hbox{MPP-2005-130} 
%\hbox{hep-ph/0511102}  
}}

\title{\phantom{x}\vspace{0.5cm} 
Renormalization Group Analysis in NRQCD for Colored Scalars
\vspace{1.0cm} }

\author{Andr\'e~H.~Hoang and Pedro~Ruiz-Femen\'\i a \vspace{0.5cm}}
\affiliation{Max-Planck-Institut f\"ur Physik\\
(Werner-Heisenberg-Institut), \\
F\"ohringer Ring 6,\\
80805 M\"unchen, Germany\vspace{1cm}
\footnote{Electronic address: ahoang@mppmu.mpg.de, ruizfeme@mppmu.mpg.de}\vspace{1cm}}

%\date{\today\\ \vspace{1cm} }

%%%%%%%%%%%%%%%%%%%%%%%%%%%%%%%%%%%%%%%%%%
\begin{abstract}
\vspace{0.5cm}
\setlength\baselineskip{18pt}
The vNRQCD Lagrangian for colored heavy scalar fields in the fundamental
representation of QCD and the renormalization group analysis of the
corresponding operators are presented. The results are an important ingredient
for renormalization group improved computations of scalar-antiscalar bound
state energies and production rates at next-to-next-to-leading-logarithmic
(NNLL) order. 
\end{abstract}
% \pacs{12.39.Hg,11.10.St,12.38.Bx}
\maketitle

%%%%%%%%%%%%%%%%%%%%%%%%%%%%%%%%%%%%%%%%%%

% \tighten
\newpage
%%%%%%%%%%%%%%%%%%%%%%%%%%%%%%%%%%%%%%%%%%
%Main body of the paper
%\setlength\baselineskip{15pt}

%
%
%
\section{Introduction}
\label{sectionintroduction}

Many models of supersymmetry (SUSY) breaking predict that at least one of the
SUSY partners of the top quark is sufficiently light such that pair production
is possible at a future Linear Collider with c.m.\,energies below 1~TeV. In
such a scenario threshold studies are feasible, where squark pairs are
produced with small relative velocities, and which allow for precise
measurements of squark masses, lifetimes or couplings in close analogy to
threshold measurements at the top-antitop pair
threshold~\cite{TTbarsim,synopsis,Habilitation}. If the SUSY 
partner of the gluon is (as is general expected) not much lighter than the
electroweak scale, then the low-energy  QCD dynamics of squarks is, up to the
fact that we are dealing with a colored spinless state, based on
standard QCD with interactions carried by spin-1 gluons in the adjoint
representation.

It is a special feature of pair production of heavy colored particles close to
the two-particle threshold that multi-gluon exchange leads to singular terms 
$\propto (\alpha_s/v)^n$ and $\propto (\alpha_s\ln v)^n$ in the amplitude at
$n$-loop perturbation theory, where $v$ is the relative particle velocity. Thus
for $v\sim\alpha_s$, which corresponds to a region in c.m.\,energy of roughly
$|\sqrt{s}-2m|\lsim 0.1 m$, $m$ being the heavy particle mass, the
singular terms have to be summed to all orders in $\alpha_s$. This is achieved
most efficiently within an effective field theory. For top pair threshold
production the effective theory vNRQCD (``velocity non-relativistic
QCD")~\cite{LMR,amis2,HoangStewartultra} was developed to carry out this
summation program aiming for NNLL order precision, i.e.\,accounting for 
${\cal O}(\alpha_s^2, \alpha_s v, v^2)$ QCD~\cite{hmst,hmst1,Hoang3loop} as well as 
${\cal O}(\alpha_{\rm em})$ electroweak
corrections~\cite{HoangReisser}. For squark-antisquark pair threshold
production previous studies involved leading-logarithmic (LL)
precision~\cite{Bigi,Fabiano}. Higher order 
QCD or electroweak effects were not taken into account in a systematic
manner, although it is known from the top threshold that corrections to the LL 
approximation can be substantial.

In this work we present the vNRQCD Lagrangian for a particle-antiparticle pair
of heavy non-relativistic colored scalars in the fundamental representation of
QCD. We provide matching conditions and anomalous dimensions required for
renormalization group improved computations of threshold pair production at
next-to-leading logarithmic (NLL) order and scalar-antiscalar bound state
energies at NNLL order in the non-relativistic expansion. For our presentation
we follow closely the conventions and 
notations of Refs.~\cite{LMR,HoangStewartultra}. Among the crucial elements of
the construction are that vNRQCD is obtained by a single matching
procedure at the hard scale $m$ and that the soft and the ultrasoft
renormalization scales, $\mu_S$ and $\mu_U$, are correlated according
to the non-relativistic energy-momentum relation of a heavy
particle at all times. Thus we have $\mu_U=\mu_S^2/m=m\nu^2$, where
$\nu$ is the dimensionless renormalization group scaling parameter of
vNRQCD.  (For corresponding computations in the pNRQCD
approach~\cite{pNRQCD2} see Ref.\,\cite{Pineda1}.)
For convenience, for the most
part of the paper, we call the
scalars in the fundamental representation of QCD simply ``squarks'' and the
scalar versions of vNRQCD and QCD frequently just vNRQCD and QCD, respectively.  

The outline of the paper is as follows: In Sec.~\ref{sectionbasic} we present the
terms of the scalar vNRQCD Lagrangian relevant for this work together with their
matching conditions. A special set of operators generated by ultrasoft
renormalization is discussed in Sec.~\ref{sectionbasic}B. The running
of all relevant potentials is discussed in Sec.~\ref{sectionpot}. Finally, 
Sec.~\ref{sectioncurrent} is devoted to the anomalous dimensions
of the S- and P-wave squark-antisquark production
currents. Conclusions are given in Sec.~\ref{sectionconclusion}.   

\section{Effective Theory Lagrangian} 
\label{sectionbasic}

The scalar vNRQCD effective Lagrangian is written in terms of the fields
$\psi_\bmp$ for the squark, $\chi_\bmp$ for the antisquark, 
($A_q^\mu$, $c_q$, $\varphi_q$, $\phi_q$) for soft (massless) gluons, ghosts,
quarks and squarks and ($A^\mu$, $c$, $\varphi$, $\phi$) for ultrasoft
(massless) gluons, ghosts, quarks and squarks. For the light quarks and
squarks we assume $n_f$ and $n_s$ light flavor components, respectively.
The covariant derivative is  $D^\mu=\partial^\mu+i g_u A^\mu=(D^0,-\bmD)$,
with  $D^0=\partial^0+i g_u A^0$, $\bmD=\nabla-i g_u \bmA$ and only involves
the ultrasoft gluon field and the ultrasoft gauge coupling. For the
representation of the (anti)squark fields and the soft fields the vNRQCD label
formalism is used to separate the fluctuations at the scales $mv$
(soft) and $mv^2$ (ultrasoft)~\cite{LMR}. Thus the spatial dependence of
fields refers only to ultrasoft fluctuations and, for example, for the heavy
squark field we have
$\partial^\mu\psi_\bmp(x)\sim mv^2\psi_\bmp(x)$, while the label $\bmp$ refers to
the soft three-momentum component of the squark field.

\subsection{Basic Lagrangian} 
\label{subsectionbasic}

The basic vNRQCD Lagrangian can be obtained by tree level matching to full QCD
and can be separated into ultrasoft, soft and potential components, 
${\cal L}={\cal L}_u + {\cal L}_s + {\cal L}_p$.\footnote{
We suppress throughout this paper the renormalization
constants that relate bare and renormalized quantities. Note that we use
the convention that the antiquark field describes {\it positive energy}
antisquarks. 
}
\begin{figure}
  \epsfxsize=9cm \centerline{\epsfbox{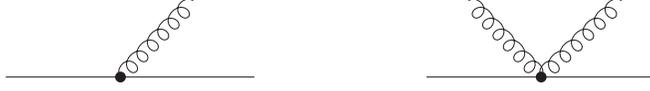}}
  \vskip 0cm
  \medskip
{\caption{Effective theory diagrams for the interaction of
a heavy squark with an ultrasoft gluon.}
\label{figusoft}}
\end{figure}
The ultrasoft piece of the effective Lagrangian describes the interactions of
ultrasoft gluons with squarks (Fig.~\ref{figusoft}) and also contains the
kinetic energy contributions for the squarks and the ultrasoft fields. It
has the form
\begin{eqnarray} 
\label{Lus}
{\cal L}_u  &=& 
\sum_{\mathbf p}\,\bigg\{
   \psi_{\bmp}^*   \bigg[ i D^0 - \frac {\left({\bf p}-i{\bf D}\right)^2}
   {2 m} +\frac{{\mathbf p}^4}{8m^3} + \ldots \bigg] \psi_{\bmp}
 + (\psi \to \chi,T\to \bar T)\,\bigg\}
\nn\\[2mm] &&
 -\,\frac{1}{4}G_u^{\mu\nu}G^u_{\mu \nu} 
 +\sum_{i=1}^{n_f} \bar\varphi_i i \slash{D} \varphi_i
 +\sum_{i=1}^{n_s} (D^\mu\phi_i)^*(D_\mu\phi_i)
 +\ldots \,,
\end{eqnarray}
where $G_u^{\mu\nu}$ is the ultrasoft field strength tensor and $m$ is
the heavy squark pole mass. The spatial dependence and the summation over
the color indices of the fields are understood here and for the rest of this paper.
To avoid double counting the interactions have to be multipole-expanded with
the scaling $\bmp\sim mv$ and $\bmD\sim mv^2$. The form of the
first term of Eq.~(\ref{Lus}) is
determined by reparametrization invariance~\cite{manohar,Manoharm3}
and does not receive any 
non-trivial renormalization factors. For the squark 
fields we use the non-relativistic normalization with 
$\psi_{\bmp}\sim\chi_{\bmp}\sim m^{3/2}$, such that the squark kinetic energy
has the usual form known from NRQCD for heavy quarks and non-relativistic
quantum mechanics. To the order we are working ${\cal L}_u$ agrees with the
ultrasoft Lagrangian for the quark case.
Likewise, the momentum and $v$ scaling of the effective theory squark
fields and effective theory Feynman diagrams is equivalent for the heavy quark
case discussed in detail in Refs.~\cite{LMR,HoangStewartultra}. Here we employ
the same power counting formalism. The factors of $\mu_U^\epsilon$ and
$\mu_S^\epsilon$ that  
appear in the effective Lagrangian in $d=4-2\epsilon$ dimensions are uniquely
determined by the requirement that the kinetic terms in the vNRQCD action are of
order $v^0$~\cite{LMR,amis2,HoangStewartultra}. To be specific, for the squark
and gluons fields this 
gives the scaling $\psi_{\bf p}\sim (mv)^{3/2-\epsilon}$, $A^\mu\sim 
(mv^2)^{1-\epsilon}$, and $A_q^\mu\sim (mv)^{1-\epsilon}$. For the covariant
derivative this leads to the form $D^\mu=\partial^\mu+i \mu_U^\epsilon g_u
A^\mu$ in dimensional regularization.

\begin{figure}
  \epsfxsize=13cm \centerline{\epsfbox{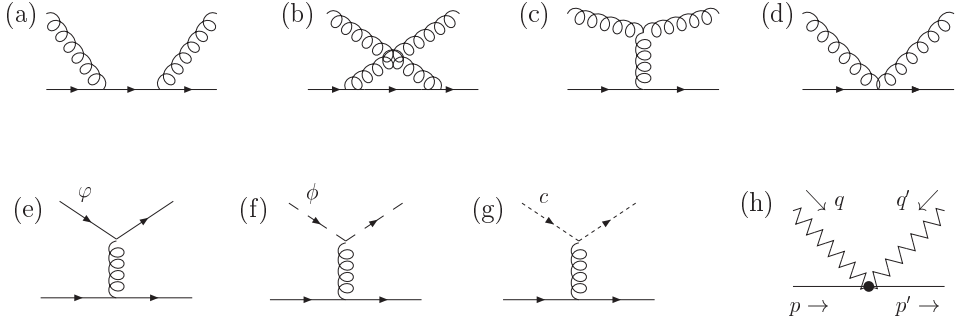}}
  \vskip 0cm
  \medskip
{\caption{The Compton scattering diagrams a-g in the full theory generate the
    effective theory soft gluon coupling h. The diagrams e, f and g describe
    interactions with soft quarks and squarks and ghosts, respectively. Soft
    gluons, ghosts, quarks and squarks in the effective theory are
    collectively depicted by zigzagged lines.}
\label{fig_softmat} }
\end{figure}

The soft Lagrangian describes interactions of soft (massless) gluons, ghosts,
quarks or squarks with the heavy squarks and has terms
\begin{eqnarray} 
\label{Ls}
{\cal L}_s &=& 
  \sum_q \Big\{
  -\frac{1}{4}G_s^{\mu\nu} G^s_{\mu \nu}
  + \bar c_q\: q^2\: c_q
  + \sum_{i=1}^{n_f}\bar\varphi_{i,q}\: \slash{q}\: \varphi_{i,q} 
  + \sum_{i=1}^{n_s}\phi_{i,q}^*\: q^2\: \phi_{i,q} \Big\}
  %+ \sum_{p} \abs{p^\mu A^\nu_p - p^\nu A^\mu_p}^2 + \ldots 
  \\
  &&
  - g_s^2 \mu_S^{2\epsilon}\! 
  \sum_{{\bmp},{\bmp^\prime},q,q^\prime,\sigma} \bigg\{ 
  \frac{1}{2}\, \psi_{\bmp^\prime}^*
  [A^\mu_{q^\prime},A^\nu_{q}] U_{\mu\nu}^{(\sigma)} \psi_{\bmp}
 + \frac{1}{2}\,
 \psi_{\bmp^\prime}^* \{A^\mu_{q^\prime},A^\nu_{q}\} W_{\mu\nu}^{(\sigma)}
 \psi_{\bmp}
 %+ (\psi \to \chi) + \ldots\, \bigg\} 
 \nn \\[2mm]\
 &&+ {\psip {p^\prime}^*}\: [\bar c_{q'}, c_q] Y^{(\sigma)}\:
 {\psip p}
 % \qquad\qquad\qquad\quad\
 + \sum_{i=1}^{n_f}
 ({\psip {p^\prime}^*}\: T^B Z_\mu^{(\sigma)}\:
 {\psip p} ) \:(\bar \varphi_{i,q'} \gamma^{\,\mu} T^B \varphi_{i,q}) 
 \nn \\[2mm]\
  &&+ \sum_{i=1}^{n_s}
  ({\psip {p^\prime}^*}\: T^B X_\mu^{(\sigma)}\:
 {\psip p} ) \:( \phi_{i,q'}^* \,(q^\mu+q^{\prime\mu})  T^B \phi_{i,q}) 
  \bigg\}
 + (\psi_{\bmp^\prime}^* T\psi_{\bmp} \to \chi_{\bmp^\prime}^* \bar T\chi_{\bmp}) \,,\nn 
\end{eqnarray}
where $G_s^{\mu\nu}$ is the soft gluon field strength tensor and $g_s=g_s(\mu_S)$
($\mu_S=m\nu$) is the soft QCD coupling. 
We also allow for $n_s$ massless scalar flavor fields $\phi_q$.
The theory graph describing the soft interactions is shown in
Fig.~\ref{fig_softmat}h. 
One-loop time-ordered products of two soft vertices with $\sigma$ and
$\sigma^\prime$ in a four-squark matrix element contribute to scattering
diagrams at order $v^{\sigma+\sigma^\prime-1}$. 
The tensors $U_{\mu\nu}^{(\sigma)}$,
$W_{\mu\nu}^{(\sigma)}$, $X_\mu^{(\sigma)}$, $Y^{(\sigma)}$, and
$Z_\mu^{(\sigma)}$ depend on the labels
${\bf p'}, {\bf p}, q, q'$ and can be obtained from matching to the full
theory diagrams in Fig.~\ref{fig_softmat}a-g.
Their explicit form in Feynman gauge and only employing the soft gluon momentum 
label $q$ is
\begin{eqnarray}  
\label{softr}
 U^{(0)}_{00} &=&  \frac{1}{q^0}\,,\quad
 U^{(0)}_{0i}  = -\frac{(2 {\bf p'}-2{\bf p-q})^i}{\ppp}\,, \quad
 U^{(0)}_{i0}  = -\frac{{\bf (p-p'-q)}^i}{\ppp}\,, \quad
 U^{(0)}_{ij}  = \frac{-2 q^0 \delta^{ij} }{\ppp} \nn \,,\\[5pt]
 U^{(1)}_{00} &=& \frac{ {\bf (p'+p)\cdot q}}{2m (q^0)^2} -
    \frac{ {\bf (p'+p)\cdot q}}{m \ppp} +{ ({\bf p'}\,^2-{\bf p}^2) \over
    2 m \ppp } \,,\nn \\[5pt]
 U^{(1)}_{0i} &=& -\frac{{\bf (p+p')}^i}{2m q^0}  + \frac{q^0 {\bf (p+p')}^i}
   {2m \ppp}  \,,
   \nn \\[5pt]
 U^{(1)}_{i0} &=& -\frac{{\bf (p+p')}^i}{2m q^0}  + \frac{q^0{\bf (p+p')}^i}{2m \ppp}
   \,,
  \nn \\[5pt]
 U^{(1)}_{ij} &=&  [ 2 \delta^{ij}
  {\bf q}^m \!+\! \delta^{im} ( {\bf \two p'\!-\! \two p\!-\!q)}^j
  \!+\! \delta^{jm} {\bf (p\!-\!p'\!-\!q)}^i]  \,
   \frac{{\bf  (p\!+\!p')}^m }  {2m ({\bf p'-p})^2}   \,, \nn\\[5pt]
 U^{(2)}_{00} &=& -\frac{c_D \ppp}{8 m^2 q^0}+
  \frac{({\bf p\cdot q})^2+({\bf p'\cdot q})^2}{2m^2(q^0)^3}
  + \frac{(1-c_D) {\mathbf q}^2}{4 m^2 q^0}\nn\\[5pt] 
   &&  +\frac{(2-c_D)
  {\bf (p-p')\cdot q}}{4m^2 q^0}\,,\nn \\[5pt]
 U^{(2)}_{0i} &=& - \frac{[ {\bf p\cdot q}\,{\bf (\two p+q)}^i
  +{\bf p'\cdot q}\: {\bf (\two p'-q)}^i\, ]}{ 4 m^2 (q^0)^2 }  
  +\frac{(c_D-1)({\bf p-p'+q})^i}{4 m^2}  \nn\\
  && +\frac{c_D({\bf 2p'-2p-q})^i}{8 m^2} 
  + \frac{ ({\bf p'}\,^2-{\bf p}^2) }{ 4 m^2 ({\bf p'-p})^2 } 
   ({\bf p}+{\bf p'})^i \nn\\[5pt]
  && - \frac{{\bf (2 p'-   2 p-q)}^i\,({\bf p'}\,^2-{\bf p}^2)^2}{4 m^2 ({\bf p'-p})^4}   
  \,,   \nn\\[5pt]
 U^{(2)}_{i0} &=& - \frac{ [{\bf p\cdot q}\,{\bf (p+p'+q)}^i +
  {\bf p'\cdot q}\: {\bf (p+p'-q)}^i\, ] }{ 4 m^2 (q^0)^2 }
+\frac{ (c_D-1){\bf q}^i}{4 m^2}    \nn \\[5pt]
  && \ +\frac{ c_D({\bf p-p'- q})^i}{8 m^2} 
  - \frac{({\bf p'}\,^2-{\bf p}^2) }{ 2 m^2 ({\bf p'-p})^2 } 
   ({\bf p}+{\bf p'})^i \nn\\[5pt]
  &&  - \frac{{\bf (p-
   p'-q)}^i\,({\bf p'}\,^2-{\bf p}^2)^2 }{4 m^2 ({\bf p'-p})^4}
   \,, \nn\\[5pt]
 U^{(2)}_{ij} &=& \frac{{\bf (p+p')}^i {\bf (p+p')}^j}{4 m^2 q^0} 
   + \frac{ {\bf q}^i({\bf p-p'+q})^j}{4 m^2 q^0}  
   - \frac{ q^0 \delta^{ij}
  ({\bf p'}\,^2-{\bf p}^2)^2 }{ 2 m^2 ({\bf p'-p})^4 } \,, \nn \\[5pt] 
 W^{(0)}_{\mu\nu} &=& 0 \,,\nn\\[5pt] 
 W^{(1)}_{00} &=& \frac{1}{2m} +\frac{{\bf (p-p')\cdot q}}{2m (q^0)^2}
   \,,\quad
 W^{(1)}_{0i}  = -\frac{{\bf (p-p'+q)}^i}{2m q^0} \,,\quad
 W^{(1)}_{i0}  = \frac{-{\bf q}^i}{2m q^0} \,,\quad
 W^{(1)}_{ij}  = \frac{\delta^{ij}}{2m} \,, 
 %\nn \\[5pt]
\end{eqnarray}
\begin{eqnarray}  
 Y^{(0)} &=&   \frac{-q^0}{\ppp} \,,\quad
 Y^{(1)} = \frac{ {\bf q \cdot (p+p')}}{2m \ppp }  \,, \quad \nn\\[5pt]
 Y^{(2)} &=& \frac{c_D q^0}{8m^2}
    \,,\nn \\
 Z^{(0)}_0 &=& \frac{1}{\ppp} \,,\quad  Z^{(0)}_i =0 \,, \quad
 Z^{(1)}_0 = 0 \,, \quad
 Z^{(1)}_i =  \frac{ -({\bf p+p'})^i }
   {2m \ppp } \,,\quad \nn\\[5pt]
 Z^{(2)}_0 &=&-\frac{c_D}{8m^2}\,
  \,,\quad
 Z^{(2)}_i = 0\nn\\[5pt]
 X^{(\sigma)}_{\mu} &=& Z^{(\sigma)}_{\mu} \,, \quad \sigma=0,1,2\,     \,.  
\label{soft_tensors}
\end{eqnarray}
Note that for completeness we also display terms proportional to  
${\bf p'}^2-{\bf p}^2$. We emphasize that we do not use these terms  
in the definition of the soft Feynman rules. In time-ordered 
products of two soft vertices these terms vanish for on-shell scattering
amplitudes. As can be checked from Eq.~(\ref{Ls}),
the tensors involving soft gluons satisfy the relations
$$
U_{\mu\nu}(q)= -U_{\nu\mu}(q') \quad,\quad 
W_{\mu\nu}(q)= W_{\nu\mu}(q') \quad,\quad q'=-q+p'-p \,,
$$
for on-shell external momenta and up to terms proportional to (${\bf
  p'}\,^2-{\bf p}^2$). The expressions for $\sigma=0$ agree exactly
with the corresponding results for the heavy quark
case~\cite{amis2,amis}, and the expressions for $\sigma=1$ also agree
up to the spin-dependent contributions that are absent for
scalars. This is related to the fact that the spin independent part of
the ${\cal O}(1/m)$ HQET (heavy quark effective theory) Lagrangian
agrees with the effective Lagrangian for a single heavy  squark, for
simplicity called HSET (heavy scalar effective theory) in the
following. This is because the spin 
independent parts only contain kinetic energy contributions which are
fixed by reparametrization invariance~\cite{manohar,Manoharm3} and do
not pick up non-trivial renormalization factors.
Since ultrasoft corrections are only accounted for in the
renormalization of time-ordered products of at least two soft
vertices describing squark-antisquark scattering,
the coefficients in ${\cal L}_s$ are 
only renormalized due to soft interactions~\cite{amis,HoangStewartultra}.  
In analogy to the quark case the structure of the resulting soft divergences 
is identical to the one in HSET. Thus one can determine the
leading-logarithmic (LL) soft running of the coefficients in  ${\cal L}_s$ by
scaling the HSET Lagrangian at order $1/m^2$, 
\begin{eqnarray}
  {\cal L}_{\rm HSET} = \psi^* \bigg\{ i D^0 +
    { {\bf D}^2 \over 2 m} +
    c_D g\,\frac{[ \bmD, \bmE ]}{8 m^2} 
   +{\cal O}\Big(\frac{1}{m^3}\Big)\,
   \bigg\} \psi 
\,,
\label{LHSET}
\end{eqnarray}
to $\mu=m \nu$, and then by matching the soft vNRQCD vertices in
${\cal L}_s$ to the corresponding HSET diagrams. This leads to the dependence
on $g_s$ and $c_D$ shown in Eqs.~(\ref{Ls}) and (\ref{softr}). Note that in
Eq.~(\ref{LHSET}) the covariant derivatives refer to soft gluon
interactions. For the solution for the coefficient $c_D(\nu)$ for
heavy scalars we find 
\begin{eqnarray}
\label{cD}
  c_D(\nu) &=& \bigg(\frac{20}{13}+\frac{32 C_F}{13 C_A}\bigg)
  \bigg[ 1 - z^{-13C_A/(6\beta_0)} \bigg] \,,
\end{eqnarray}
with
\begin{eqnarray}
z & \equiv & \frac{\alpha_s(m\nu)}{\alpha_s(m)}
\,,
\nn\\[2mm] 
\beta_0 & = & \frac{11}{3}\, C_A 
- \frac{4 n_f + n_s}{3}\,T_F
\,
\end{eqnarray}
and the group theoretical factors $C_A=3$, $C_F=4/3$, $T_F=1/2$.
The term $\beta_0$ is the one-loop QCD beta-function that also accounts for
the running due to the light squark flavors. Note that the coefficient $c_D$
in HSET vanishes at the hard scale, $c_D(1)=0$ as shown in the
appendix~\ref{HSETconstruction} 
where the tree level HSET Lagrangian is derived. 
The evolution of $c_D$ can be deduced from the results given in
Ref.~\cite{Korner,Bauer1} and differs from the corresponding coefficient in HQET
because for the heavy 
scalar field the mixing to spin-dependent operators does not occur. 
The relation of the tensors in Eqs.~(\ref{softr}) to the HSET Lagrangian 
also explains (except for the dependence on the
spin-dependent 
HQET coefficients $c_F$ and $c_S$ which are absent in the scalar case)
the close resemblance 
of the $\sigma=2$ tensors to the quark results obtained
in Ref.~\cite{amis2,amis}. 
We note that in Eqs.~(\ref{softr}) all $i\epsilon$ prescriptions in the $1/q^0$
terms originating from integrating out the intermediate static squark are
dropped. This prevents pinch-singularities from the $1/q^0$ term in one-loop 
diagrams related to time-ordered products of soft vertices. 

\begin{figure}
  \epsfxsize=16cm \centerline{\epsfbox{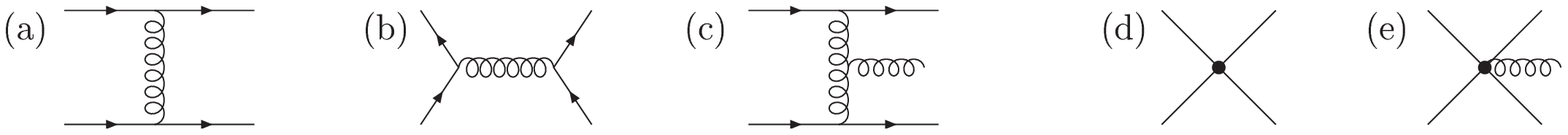}}
  \vskip 0cm
  \medskip
{\caption{(a-c) QCD tree level diagrams relevant for the LL matching
conditions of the potentials; (d,e) effective theory Feynman diagrams for the
potential interactions.
}
\label{fig_pot}}
\end{figure}

The potential Lagrangian describes potential-like squark-antisquark scattering
interactions and has terms
\begin{eqnarray}
{\cal L}_p &=& -\sum_{\bmp,\bmp^\prime}\mu_S^{2\epsilon}\, V({\bmp,\bmp^\prime})\,
   \psi_{\bmp^\prime}^* \psi_{\bmp}
   \chi_{-\bmp^\prime}^* \chi_{-\bmp}\nn \\[3mm]
 && +\sum_{\bmp,\bmp^\prime}
    \mu_S^{2\epsilon} F_j^{ABC}({\bf p},{\bf p'})(g_u\mu_U^\epsilon {\bmA}^C_j) 
   \Big[\psi_{\bmp^\prime}^* T^A \psi_{\bmp} 
   \chi_{-\bmp^\prime}^* \bar T^B \chi_{-\bmp} \Big]  + \ldots \,,
\label{Lp}
\end{eqnarray}  
where
\begin{eqnarray}
 V({\bmp},{\bmp^\prime})  &=&   (T^A \otimes \bar T^A) \bigg[
 \frac{{\cal V}_c^{(T)}}{\bmk^2}
 + \frac{{\cal V}_k^{(T)}\pi^2}{m|{\bmk}|}
 + \frac{{\cal V}_r^{(T)}({\bmp^2 + \bmp^{\prime 2}})}{2 m^2 \bmk^2}
 + \frac{{\cal V}_2^{(T)}}{m^2}
 + \ldots \bigg] \nn \\[2mm] 
 && 
 + (1\otimes 1)\bigg[
 \frac{{\cal V}_c^{(1)}}{\bmk^2}
 + \frac{{\cal V}_k^{(1)}\pi^2}{m|{\bmk}|}
 + \frac{{\cal V}_2^{(1)}}{m^2} +\ldots \bigg]
\,, \nonumber \\[3mm]
F^{ABC}_j({\bmp},{\bmp^\prime}) & = & 
\frac{2i{\cal V}_c^{(T)}\bmk_j }{\bmk^4}  f^{ABC} \,,
\label{vNRQCDpotential}
\end{eqnarray}
and $\bmk\equiv\bmp'-\bmp$. 
The terms with the coefficients ${\cal V}_c^{(1,T)}$ are the Coulomb
potential that contributes at order $\alpha_s v^{-1}$ to scattering diagrams. 
The terms with the coefficients ${\cal V}_{2}^{(1,T)}$ and 
${\cal V}_{r}^{(T)}$ contribute at order $\alpha_s v$ and are the scalar
analog of the Breit-Fermi potentials known 
from QED. The terms with the coefficients ${\cal V}_k^{(1,T)}$ are only
generated by one-loop diagrams~\cite{HoangStewartultra} and contribute at
order $\alpha_s^2 v^0$. For scattering of squark-antisquark pairs in the color
singlet state only the linear combination of coefficients 
${\cal V}_i^{(s)}=-C_F{\cal V}_i^{(T)}+{\cal V}_i^{(1)}$ is relevant.
Graphically the first and second terms in
Eq.~(\ref{Lp}) are depicted in Fig~\ref{fig_pot}d and e, respectively.
Matching to the full theory Born diagrams in Fig.~\ref{fig_pot}a,b leads to
the LL matching conditions at $\nu=1$,
\begin{eqnarray} 
 \label{bc}
&&
 {\cal V}_c^{(T)}(1) = 4 \pi \alpha_s(m)\,, \qquad
 {\cal V}_r^{(T)}(1) = 4 \pi \alpha_s(m)\,, \qquad
 {\cal V}_2^{(T)}(1) = - \pi \alpha_s(m) \,,
\nn\\[2mm] & &
 {\cal V}_c^{(1)}(1) = 0\,,\hspace{2.2cm}
 {\cal V}_2^{(1)}(1) = 0
\,.
\end{eqnarray}
These matching conditions are consistent with the results given earlier in
Ref.~\cite{Gupta}. Note that the matching condition for 
${\cal V}_2^{(T)}$ differs from the corresponding result in the quark
case. The annihilation diagram in Fig.~\ref{fig_pot}b does not
contribute to scattering for a squark-antisquark pair in a color
singlet state. Moreover, the annihilation diagram contributes at order
$\alpha_s v^3$ because a squark-antisquark pair that annihilates into
a gluon is in a P-wave state. Although the results will not be needed
at the order we are interested in, we give them here for completeness: 
\begin{eqnarray}
 V_{a}({\bmp},{\bmp^\prime})   &=&   (T^A \otimes \bar T^A) \bigg[
 \,\frac{{\cal V}_{a}^{(T)}\,\bmp^\prime.\bmp}{m^4}
 + \ldots \bigg] 
 + (1\otimes 1)\bigg[
 \,\frac{{\cal V}_{a}^{(1)}\,\bmp^\prime.\bmp}{m^4}
  +\ldots \bigg]
  \,,
\label{annihilation}
\end{eqnarray}
where
\begin{eqnarray}  
&&
 {\cal V}_{a}^{(T)} \,=\, \frac{1}{C_A}\pi \alpha_s(m)
 \,,\qquad
 {\cal V}_{a}^{(1)} = \frac{C_F}{C_A}\, \pi \alpha_s(m)
\,.
\end{eqnarray}
The matching conditions for the coefficients ${\cal V}_k^{(1,T)}$ arise from
the terms in the one-loop full theory scattering diagrams contributing at
order $\alpha_s^2 v^0$. Since in the effective theory all one-loop
contributions at order $\alpha_s^2 v^0$ from time-ordered products of soft
vertices with $\sigma=0$ and $\sigma^\prime=1$ vanish the matching conditions
are just equal to the corresponding full theory contributions. The latter can
be obtained using the threshold expansion and read 
\begin{eqnarray}  
\label{1overk} 
&&
 {\cal V}_k^{(T)}(1) = \alpha_s^2(m)\left( \frac{7C_A}{8}-\frac{C_d}{8}\right)
\,, \qquad{\cal V}_k^{(1)}(1) = \alpha_s^2(m) \frac{C_1}{2} \,,
\end{eqnarray}
where
\begin{eqnarray}
 C_d=8C_F-3C_A \,, \qquad C_1=\frac{1}{2}\,C_F C_A-C_F^2\,. 
\end{eqnarray}
The results agree with the corresponding matching conditions for the quark
case~\cite{HoangStewartultra,amis2}. Similar to the agreement for the soft
$\sigma=0,1$ vertices this feature can again be traced back to the equivalence
of the HQET and HSET effective actions mentioned above.

\subsection{Operators Generated by Ultrasoft Renormalization} 
\label{subsectionultra}

In the potential and soft components of the vNRQCD Lagrangian there are a
number 
of operators with Wilson coefficients that have vanishing matching conditions
at $\nu=1$, but which become non-zero for $\nu<1$ due to ultrasoft
renormalization~\cite{HoangStewartultra}. Since the structure of these
operators only depends 
on the form of the ultrasoft Lagrangian ${\cal L}_u$ and on operators in the
soft and potential Lagrangians ${\cal L}_{s,p}$ that can contribute to
scattering terms at order $v^{-1}$, they are in complete analogy to
the corresponding 
operators in the quark case already discussed in detail in  
Ref.~\cite{HoangStewartultra}, 
apart from the operators involving soft squarks which were not considered
there. For completeness we briefly discuss the full set of these
operators relevant to the order we are interested in. 

\begin{figure}[t!]
\centerline{ 
\includegraphics[width=4.8in]{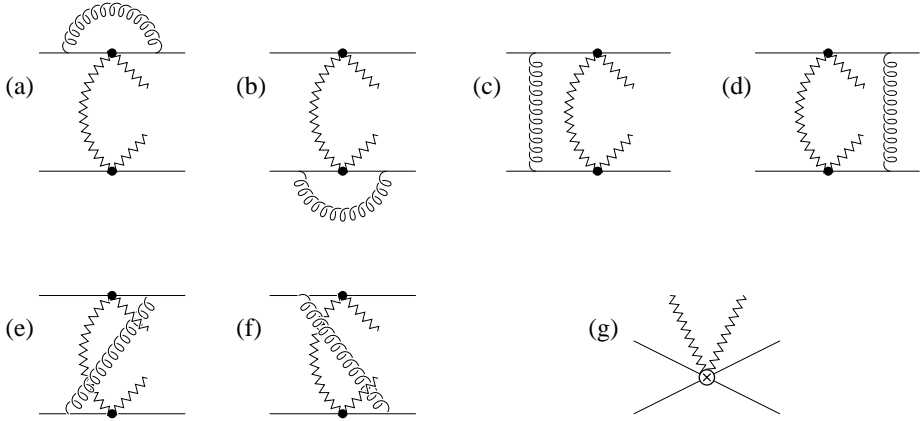}
}
\vspace{0.5cm}
\caption{(a-f) Ultrasoft diagrams  with $\bmp.\bmA$
vertices for the computation of the LL anomalous dimension of the
operators ${\cal O}_{2i}^{(2),(1,T)}$ (wave function renormalization
graphs are understood); (g) graphical representation of operators  
${\cal O}_{2i}^{(2),(1,T)}$, where $i=a,b,c$.    }
\label{figDeltaLsoftusoft}
\end{figure}

In the soft sector we have to include 6-field operators describing
squark-antisquark scattering with two additional soft massless gluons, ghosts, 
quarks or squarks. The additional terms in the soft Lagrangian have the form
\begin{eqnarray}
\Delta {\cal L}_s & = &
 C_{2a}^{(2)}{\cal O}_{2a}^{(2),(1)} +
 C_{2b}^{(2)}{\cal O}_{2b}^{(2),(T)} +
 C_{2c}^{(2)}{\cal O}_{2c}^{(2),(T)}
\,,
\label{Ls_add}
\end{eqnarray}
where
\begin{eqnarray} 
\label{O2abc}
 {\cal O}_{2a}^{(2),(1)} &=&  \frac{\bmk^2}{m^2}\, \sum_i {\cal O}_{2i}^{(0),(1)}
   \,,\quad
 {\cal O}_{2b}^{(2),(T)} \,=\,  \frac{\bmk^2}{m^2}\, \sum_i {\cal O}_{2i}^{(0),(T)}
   \,,\quad \nn\\
 {\cal O}_{2c}^{(2),(T)} &=&  \frac{(\bmp^2+\bmp^{\prime\,2})}{m^2}\, 
   \sum_i {\cal O}_{2i}^{(0),(T)} \,,
\end{eqnarray}
and the sums are related to the operators
\begin{eqnarray} \label{6q}
  {\cal O}_{2\varphi}^{(0)} &=& { g_s^4\,\mu_S^{4\epsilon}}\:  \:
   (\psi_{\bmp^\prime}^*\: \Gamma^{(0)}_{\varphi,\psi}\: 
   \psi_{\bmp}) \:
   (\chi_{-\bmp^\prime}^*\: \Gamma^{(0)}_{\varphi,\chi}\: 
   \chi_{-\bmp})
   \ (\overline\varphi_{q}\: \Gamma_\varphi^{(0)}\: \varphi_{q})\,,\nn\\\
 {\cal O}_{2\phi}^{(0)}  &=& { g_s^4\,\mu_S^{4\epsilon}}\:  \:
   (\psi_{\bmp^\prime}^*\: \Gamma^{(0)}_{\phi,\psi} \: \psi_{\bmp})  \:
   (\chi_{-\bmp^\prime}^*\: \Gamma^{(0)}_{\phi,\chi} \: \chi_{-\bmp})
   \ (\phi_{q}^*\: \Gamma_\phi^{(0)}\: \phi_{q}) \,, \nn\\
 {\cal O}_{2A}^{(0)}  &=& { g_s^4\,\mu_S^{4\epsilon}}\:  \:
   (\psi_{\bmp^\prime}^*\: \Gamma^{(0)}_{A,\psi} \: \psi_{\bmp})  \:
   (\chi_{-\bmp^\prime}^*\: \Gamma^{(0)}_{A,\chi} \: \chi_{-\bmp})
   \ (A^\mu_{-q}\: \Gamma_{A,\mu\nu}^{(0)}\: A^\nu_{q}) \,,\nn\\\
 {\cal O}_{2c}^{(0)}  &=& { g_s^4\,\mu_S^{4\epsilon}}\:  \:
   (\psi_{\bmp^\prime}^*\: \Gamma^{(0)}_{c,\psi} \: \psi_{\bmp})  \:
   (\chi_{-\bmp^\prime}^*\: \Gamma^{(0)}_{c,\chi} \: \chi_{-\bmp})
   \ (\bar c_{q}\: \Gamma_c^{(0)}\: c_{q})  \,,
\end{eqnarray}
referring to soft quarks, squarks, gluons and ghosts, respectively. 
Note that in Eq.~(\ref{Ls_add}) we have suppressed all sums over soft
labels and flavor indices.
The operators  ${\cal O}_{2i}^{(2)}$ are represented graphically by the
diagram in Fig.~\ref{figDeltaLsoftusoft}g. The matrices $\Gamma_i$ are
functions of the heavy 
squark and soft labels and have the form
\begin{eqnarray}
\Gamma_\varphi^{(0),(1)} &=& \Gamma_\varphi^{(0),(T)}
= 
\frac{1}{2}\bigg[\,
\frac{(2q^0\gamma^0+{\bf k}\cdot{\mbox{\boldmath$\gamma$}})}
 {{\bf k}^2+2 {\bf k}\cdot {\bf q}} +
 \frac{(2q^0\gamma^0-{\bf k}\cdot{\mbox{\boldmath $\gamma$}})}
 {{\bf k}^2-2 {\bf k}\cdot {\bf q}}
\,\bigg]\,\Big(Z_0^{(0)}\Big)^2
\,,
\nonumber\\[2mm]
\Gamma_{\varphi,\psi}^{(0),(T)} &=& T^A\,,\quad 
\Gamma_{\varphi,\chi}^{(0),(T)} = \bar T^A\,,\quad
\Gamma_{\varphi,\psi}^{(0),(1)} = 1\,,\quad
\Gamma_{\varphi,\chi}^{(0),(1)} = 1\,,
\nonumber\\[2mm]
\Gamma_\phi^{(0),(1)} &=& \Gamma_\phi^{(0),(T)}
= 
2\bigg[\,
\frac{(q^0)^2}{{\bf k}^2+2 {\bf k}\cdot {\bf q}} +
\frac{(q^0)^2}{{\bf k}^2-2 {\bf k}\cdot {\bf q}}
\,\bigg]\,\Big(X_0^{(0)}\Big)^2
\,,
\nonumber\\[2mm]
\Gamma_{\phi,\psi}^{(0),(T)} &=& T^A\,,\quad 
\Gamma_{\phi,\chi}^{(0),(T)} = \bar T^A\,,\quad
\Gamma_{\phi,\psi}^{(0),(1)} = 1\,,\quad
\Gamma_{\phi,\chi}^{(0),(1)} = 1\,,
\nonumber\\[2mm]
\Gamma_{A,\mu\nu}^{(0),(T)} &=& \Gamma_{A,\mu\nu}^{(0),(1)} 
=
\frac{C_A}{2}\,\bigg\{\,\Big[\,
U^{(0)}_{\sigma\mu}(-q,\bmp,\bmp^\prime)\,
(U^{(0)})^\sigma_{\,\,\,\,\nu}(q,-\bmp,-\bmp^\prime)\,
\Big]\,
\frac{1}{{\bf k}^2+2 {\bf k}\cdot {\bf q}}
\nonumber\\&& \qquad\quad
+ \,
\Big[\,
U^{(0)}_{\sigma\nu}(q,\bmp,\bmp^\prime)\,
(U^{(0)})^\sigma_{\,\,\,\,\mu}(-q,-\bmp,-\bmp^\prime)\,
\Big]\,
\frac{1}{{\bf k}^2-2 {\bf k}\cdot {\bf q}}
\,\bigg\}
\,,
\nonumber\\[2mm]
\Gamma_{A,\psi}^{(0),(T)} &=& T^A\,,\quad 
\Gamma_{A,\chi}^{(0),(T)} = \bar T^A\,,\quad
\Gamma_{A,\psi}^{(0),(1)} = 1\,,\quad
\Gamma_{A,\chi}^{(0),(1)} = 1\,,
\nonumber\\[2mm]
\Gamma_{c,\psi}^{(0),(T)} &=& \Gamma_{c,\psi}^{(0),(1)}
= 
C_A\, \bigg[\, 
\frac{1}{{\bf k}^2+2 {\bf k}\cdot {\bf q}} +
\frac{1}{{\bf k}^2-2 {\bf k}\cdot {\bf q}}
\,\bigg]\,\Big(Y_0^{(0)}\Big)^2 \,,
\nonumber\\[2mm]
\Gamma_{c,\psi}^{(0),(T)} &=& T^A\,,\quad 
\Gamma_{c,\chi}^{(0),(T)} = \bar T^A\,,\quad
\Gamma_{c,\psi}^{(0),(1)} = 1\,,\quad
\Gamma_{c,\chi}^{(0),(1)} = 1\,. 
\label{Gam1} 
\end{eqnarray}
The color indices for the two soft particles have been
contracted in the expressions in Eq.~(\ref{Gam1}) because at the order
we are interested in we only need the operators in terms of
one-loop diagrams where the soft lines are closed up (see
Fig.~\ref{figVm2renor}h). When the soft lines are closed up the
operators contribute to scattering matrix elements at order $\alpha_s
v$. 
The Wilson coefficients run due to UV-divergences from ultrasoft gluon
loops  (and the corresponding soft pull-up terms) dressing the
time-ordered products of two soft vertices with
$\sigma=\sigma^\prime=0$, see Fig.~\ref{figDeltaLsoftusoft}. The
results have been  worked out in Ref.~\cite{HoangStewartultra} and
read 
\begin{eqnarray} 
\label{C2slnQCD}
 C_{2a}^{(2)}(\nu) & = & \frac{4C_1}{3\beta_0}
    \ln(w)
  \,,\qquad\quad
 C_{2b}^{(2)}(\nu) =  \frac{3C_A\!-\!C_d\!-\!4C_F}{3\beta_0}
    \ln(w)
   \,,\nn\\[2mm]
 C_{2c}^{(2)}(\nu) &=& \frac{-4C_A}{3\beta_0}\, 
  \ln(w)
 \,,
\label{solutionvRGEQCD}
\end{eqnarray}
where
\begin{eqnarray} \label{w}
  w = \frac{\alpha_s(m\nu^2)}{\alpha_s(m\nu)}  \,.
\end{eqnarray}

\begin{figure}[t!]
  \epsfxsize=11cm \centerline{\epsfbox{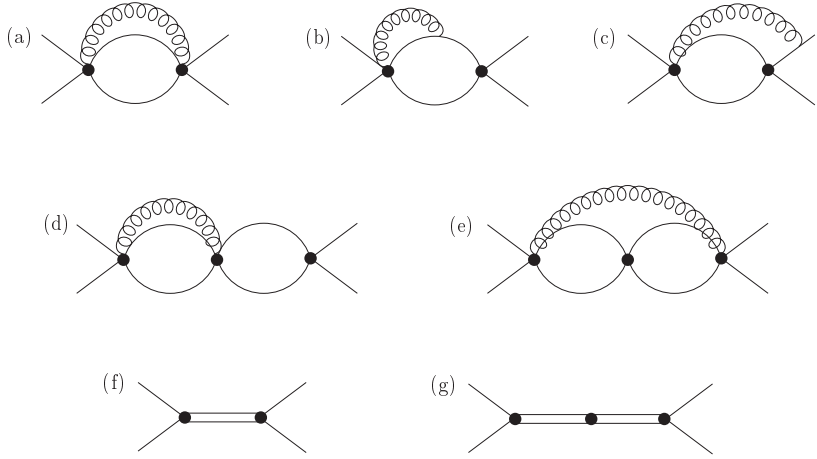}}
  \vskip 0cm
  \medskip
{\caption{(a-c) Ultrasoft graphs for the renormalization of the operators
    ${\cal O}_{k1,k2}^{(1,T)}$, (d,e) ultrasoft graphs for the renormalization of
    the operators ${\cal O}_{c1,c2,c3}^{(1,T)}$, (f,g) graphical representation
    of operators ${\cal O}_{k1,k2}^{(1,T)}$ and ${\cal O}_{c1,c2,c3}^{(1,T)}$.}
\label{figDeltaLpotusoft}}
\end{figure}

In the potential sector we have to account for additional 4-squark
operators which involve sums over intermediate 3-momentum labels. The
additional terms in the potential Lagrangian are
\begin{eqnarray}
\Delta {\cal L}_p & = &
 {\cal V}_{k1}^{(1)}{\cal O}_{k1}^{(1)} +
 {\cal V}_{k2}^{(T)}{\cal O}_{k2}^{(T)} +
%\nn\\[2mm] & &
 {\cal V}_{c1}^{(1)}{\cal O}_{c1}^{(1)} +
 {\cal V}_{c2}^{(T)}{\cal O}_{c2}^{(T)} +
 {\cal V}_{c3}^{(T)}{\cal O}_{c3}^{(T)}
\,,
\label{Lp_add}
\end{eqnarray}
where
\begin{eqnarray} 
\label{Okc}
{\cal O}_{k1}^{(1)} &=& - \frac{[\mu_S^{2\epsilon}\,{\cal V}_c^{(T)}]^2}{m}\: 
 \sum_{{\bf p,p',q}} ( f_{0} + f_{1} + 2 f_{2} )\
 \big[ \psi_{\bmp^\prime}^* \psi_{\bmp}
 \chi_{-\bmp^\prime}^* \chi_{-\bmp} \big] \,, \nn\\
{\cal O}_{k2}^{(T)} &=& - \frac{[\mu_S^{2\epsilon}\,{\cal V}_c^{(T)}]^2}{m}\: 
 \sum_{{\bf p,p',q}} ( f_{1} + f_{2} ) \ 
 \big[ \psi_{\bmp^\prime}^* T^A \psi_{\bmp}
 \chi_{-\bmp^\prime}^* \bar T^A \chi_{-\bmp} \big] \,, \nn\\
  {\cal O}_{c1}^{(1)} &=& - [\mu_S^{2\epsilon}\,{\cal V}_c^{(T)}]^3
   \sum_{{\bf p,p',q,q'}}   (2h_0-h_1)\:
    \big[ \psi_{\bmp^\prime}^* \psi_{\bmp}
   \chi_{-\bmp^\prime}^* \chi_{-\bmp} \big] \,,
 \nn\\
   {\cal O}_{c2}^{(T)} &=& -[\mu_S^{2\epsilon}\,{\cal V}_c^{(T)}]^3 
   \sum_{{\bf p,p',q,q'}} h_0\: \big[ \psi_{\bmp^\prime}^* T^A \psi_{\bmp}
   \chi_{-\bmp^\prime}^* \bar T^A \chi_{-\bmp} \big] \,,
 \nn\\
   {\cal O}_{c3}^{(T)} &=& -[\mu_S^{2\epsilon}\,{\cal V}_c^{(T)}]^3
   \sum_{{\bf p,p',q,q'}}  h_1 \: \big[\psi_{\bmp^\prime}^* T^A \psi_{\bmp}
   \chi_{-\bmp^\prime}^* \bar T^A \chi_{-\bmp} \big]
 \,,
\end{eqnarray}
and
\begin{eqnarray}
 && f_0 = \frac{\bmp^\prime\cdot (\bmq-\bmp)}{(\bmq-\bmp)^4\,
  (\bmq-\bmp^\prime)^2} 
  + (\bmp \leftrightarrow \bmp^\prime)\,, \qquad
 f_1 = \frac{\bmq \cdot (\bmq-\bmp)}{(\bmq-\bmp)^4\,(\bmq-\bmp^\prime)^2} 
  + (\bmp \leftrightarrow \bmp^\prime)\,, \nn\\
 && f_2 = \frac{(\bmq-\bmp^\prime)\cdot (\bmq-\bmp)}
  {(\bmq-\bmp)^4\,(\bmq-\bmp^\prime)^4}\: (\bmq^2-\bmp^{\prime\,2}/2-\bmp^2/2) 
  \,,\nn\\ &&
  h_0 = \frac{(\bmq^\prime-\bmp^\prime)\cdot (\bmq-\bmp)}
   {(\bmq-\bmp)^4(\bmq-\bmq^\prime)^2(\bmq^\prime-\bmp^\prime)^4} 
   \,, \quad \nn\\ &&
  h_1 = \frac{(\bmq-\bmq^\prime)\cdot (\bmq-\bmp)}
   {(\bmq-\bmp)^4(\bmq-\bmq^\prime)^4(\bmq^\prime-\bmp^\prime)^2} 
   + (\bmp \leftrightarrow \bmp^\prime,\bmq\leftrightarrow \bmq^\prime )\,.
\end{eqnarray}
The operators are needed due to UV-divergences in the
diagrams in Figs.~\ref{figDeltaLpotusoft}a-e
after the ultrasoft integrations in the ultrasoft {\it and} the potential
loops have been performed, but before the remaining sums over the
squark-antisquark soft 3-momenta labels are carried out. 
The operators ${\cal O}_{ki}^{(1,T)}$ arise from the order $\alpha_s^3 v^0$
diagrams Figs.~\ref{figDeltaLpotusoft}a-c and are graphically depicted by 
Fig.~\ref{figDeltaLpotusoft}f. The operators ${\cal O}_{ci}^{(1,T)}$ arise
from the order $\alpha_s^4 v^{-1}$ diagrams
Figs.~\ref{figDeltaLpotusoft}d,e and are graphically depicted by
Fig.~\ref{figDeltaLpotusoft}g.  The Wilson coefficients were computed in
Ref.~\cite{HoangStewartultra} and read 
\begin{eqnarray} 
\label{Vk1Vc1}
  {\cal V}_{k1}^{(1)}(\nu) &=& \frac{8C_A C_1}{3\beta_0}\: \ln(w)\,,
  \qquad\qquad\qquad\qquad
  {\cal V}_{k2}^{(T)}(\nu) = -\frac{2C_A(C_A+C_d)}{3\beta_0}\: \ln(w)\,, 
   \nn\\[3pt]
  {\cal V}_{c1}^{(1)}(\nu) &=& -\frac{C_A C_1(C_A+C_d)}{3\beta_0}\:\ln(w)\,,
    \qquad\qquad
  {\cal V}_{c2}^{(T)}(\nu) = -\frac{16C_A C_1}{3\beta_0}\: \ln(w)\,, 
   \nn\\[3pt]  
  {\cal V}_{c3}^{(T)}(\nu) &=& -\frac{8C_A}{3\beta_0}
     \bigg[ C_1 + \frac{(C_A+C_d)^2}{32} \bigg] \ln(w) \,.
\end{eqnarray}
Explicit expressions for 4-squark matrix elements of the operators 
${\cal O}_{ki}$ and ${\cal O}_{ci}$ with the sum over labels carried out in
$n=3-2\epsilon$ dimensions can be found in the appendix of
Ref.~\cite{HoangStewartultra}.

Note that it is possible to choose a scheme where the ${\cal V}_k$ potentials
in Eq.~(\ref{vNRQCDpotential}) are also written as sum operators in analogy to 
Eq.~(\ref{Lp_add}). In this scheme the UV-finite contributions in NNLL matrix
elements and the UV-divergences in NNNLL matrix elements differ. More details
on this scheme are given in Ref.~\cite{Hoang3loop}.

\section{Anomalous dimensions for Potentials} 
\label{sectionpot}

\begin{figure}[t!]
  \epsfxsize=12cm \centerline{\epsfbox{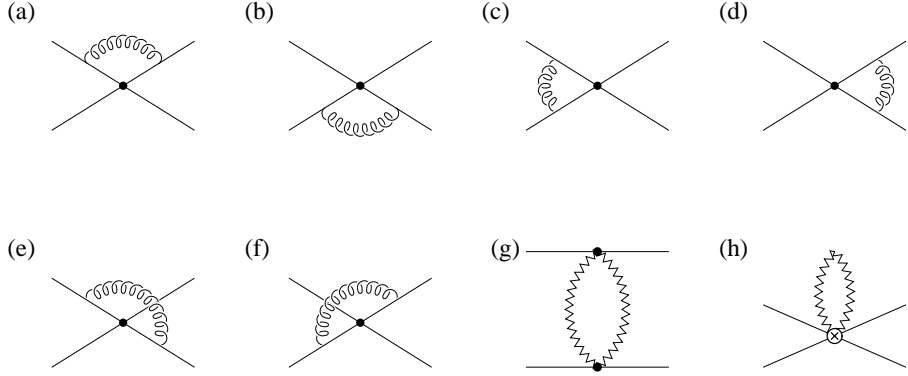}}
  \vskip 0cm
  \medskip
{\caption{Diagrams for the computation of the LL anomalous dimension of the
order $\alpha_s v$ potentials: a-f) ultrasoft gluon graphs with $\bmp.\bmA$
vertices (wave function renormalization graphs are understood), g) soft
graphs with $\sigma+\sigma^\prime=2$, h) soft graphs involving the operators 
${\cal O}_{2i}^{(2)}$.}
\label{figVm2renor} }
\end{figure}

The effective theory one-loop graphs required to determine the LL
anomalous dimension of the order $v$ potentials in
Eq.~(\ref{vNRQCDpotential}) are shown in Fig.~\ref{figVm2renor}. Note
that for ultrasoft gluon diagrams only the  $\bmp.\bmA$ couplings have
to be accounted for because the contributions from the time-like $A^0$
gluons vanish~\cite{amis}. This feature can be formally understood
from the fact that the leading order coupling of the heavy particle
with time-like gluons can be removed by field
redefinition~\cite{Korchemsky}.  
We find the anomalous dimensions 
\begin{eqnarray}  
\label{VRG}
\nu {\partial \over \partial\nu} {\cal V}_r^{(T)}(\nu)
  & = & -2 (\beta_0 + \frac{8}{3}
  C_A )\, \alpha_s^2(m\nu) + \frac{32}{3} C_A\, \alpha_s(m\nu) \alpha_s(m\nu^2)
  -4\beta_0\,C_{2c}^{(2)}(\nu)\, \alpha_s^2(m\nu)
  \,, \nn\\[5pt]
\nu \frac{\partial}{\partial\nu} {\cal V}_2^{(T)}(\nu) 
  & = &  \left(\frac{\beta_0}2 \Big[1+c_D(\nu) \Big]+
  \frac{C_A}{6}  \Big[28 - 11 c_D(\nu) \Big] -
  \frac{7C_d}{6}  \right) \, \alpha_s^2(m\nu) \nn\\[3pt] 
  & & + \,\frac{4}{3} (4C_F+C_d-3C_A)\, \alpha_s(m\nu) \alpha_s(m\nu^2) 
  -2\,\beta_0\, C_{2b}^{(2)}(\nu)\, \alpha_s^2(m\nu)
  \,,  \nn\\[3pt]
\nu \frac{\partial}{\partial\nu} {\cal V}_2^{(1)}(\nu) 
  & = & \frac{14}{3}\, C_1\, \alpha_s^2(m\nu) - \frac{16}{3}\,C_1\, 
  \alpha_s(m\nu) \alpha_s(m\nu^2)
  -2\,\beta_0\, C_{2a}^{(2)}(\nu)\, \alpha_s^2(m\nu) \,. 
\end{eqnarray}
The results for ${\cal V}_r^{(T)}$ and ${\cal V}_2^{(1)}$ agree with the
results for heavy quarks obtained in~\cite{HoangStewartultra} (and are
consistent with corresponding results in Ref.~\cite{Pineda2}), while the
result for ${\cal V}_2^{(T)}$ differs due to the absence of the spin 
dependent HQET Wilson coefficient $c_F$. Concerning the structure of
logarithms at order $\alpha_s^2$ described by Eq.~(\ref{VRG}) we
disagree with the results given in Ref.~\cite{Gupta}. Note that part
of the discrepancy is related to approximations made in
Ref.~\cite{Gupta} that are unjustified in the NRQCD $v$ power counting 
scheme. (See also Ref.~\cite{amis2} for more details.) Accounting for
the matching conditions
shown in Eqs.~(\ref{bc}) the solutions for the coefficients read
\begin{eqnarray}  
\label{coef} 
{\cal V}_r^{(T)}(\nu) 
  & = & 4\pi\,\alpha_s(m)\,z  
  - \frac{32\pi C_A}{3\beta_0}\, \alpha_s(m)\,z\,\ln(w) \,, \\[3pt]
{\cal V}_2^{(T)}(\nu) 
  & = & -\pi z \, \alpha_s(m) \nn \\[3pt] 
  & + & \frac{ \pi \left[ C_A (152C_F-35C_A) 
  -4\beta_0(5 C_A + 8 C_F) \right]}
  { 13\beta_0\, C_A }\,\alpha_s(m)\left(z - 1\right) \nn\\[3pt] 
  & + & \frac{8 \pi(3\beta_0 - 11C_A)(5C_A + 8C_F)}
  {13 C_A(6\beta_0 - 13 C_A)}\,\alpha_s(m)\,
  \left[z^{1 - 13C_A/(6\beta_0)} - 1\right] \nn\\[3pt] 
  & - &  \frac{4\pi(4C_F -2C_A)}{\beta_0}\,\alpha_s(m)\,z \ln(w) 
  \,, \nn \\[3pt]
{\cal V}_2^{(1)}(\nu) 
  & = & \frac{4\pi C_1}{\beta_0}\,\alpha_s(m)\,
  \left(1 - z\right) +\frac{16\pi C_1}{3\beta_0}\,\alpha_s(m)\,z\ln(w)
  \,.
\end{eqnarray}

The running of the coefficients ${\cal V}_{c,k}^{(1,T)}$ is determined solely
by soft diagrams and associated with the known running of the strong
coupling~\cite{HoangStewartultra}. The results for the anomalous dimensions
relevant for describing the squark-antisquark dynamics at NNLL order are 
\begin{eqnarray}  
\label{VRG2}
\nu {\partial \over \partial\nu} {\cal V}_c^{(T)}(\nu)
  & = & 
  -2  \bigg[ \beta_0 \alpha_s^2(m\nu) 
    +\beta_1 \frac{\alpha_s^3(m\nu)}{4\pi}
    +\beta_2 \frac{\alpha_s^4(m\nu)}{(4\pi)^2} \bigg] \nn\\[5pt]
\nu \frac{\partial}{ \partial\nu} {\cal V}_k^{(T)}(\nu) 
  & = & -\frac{\beta_0}{8\pi}  (7 C_A-C_d) \: {[\alpha_s(m\nu)]^3} \,, \nn \\[5pt]
\nu \frac{\partial }{ \partial\nu} {\cal V}_k^{(1)}(\nu) 
  & = & - \frac{\beta_0}{2\pi}  \,C_1\: {[\alpha_s(m\nu)]^3} \nn \,,
\end{eqnarray}
where $\beta_{0,1,2}$ are the coefficients of the QCD beta function
(in the $\overline{\rm MS}$ scheme for $\beta_2$). The anomalous dimension for
the Coulomb potential is needed at three loops because the Coulomb potential
already contributes at order $\alpha_s v^{-1}$. The results agree with the
results obtained in the quark case~\cite{HoangStewartultra}. For the running
of the Coulomb potential the agreement is obvious since 
the respective leading order interactions of heavy squarks and quarks with low
energy gluons are universal. The agreement for the order $v^0$ potential
coefficients ${\cal V}_k^{(1,T)}$ is again related to the equivalence of
the HQET and HSET effective actions at order $1/m$ discussed before.
With the matching conditions shown in Eqs.~(\ref{bc}) and (\ref{1overk}) 
the solutions read
\begin{eqnarray} 
\label{VkVc}
  {\cal V}_k^{(T)}(\nu) = \frac{(7C_A\!-\!C_d)}{8}\: \alpha_s^2(m\nu)\,,\quad
  {\cal V}_k^{(1)}(\nu) = \frac{C_1}{2}\: \alpha_s^2(m\nu)\,,\quad
  {\cal V}_c^{(T)}(\nu) = 4\pi \alpha_s^{[3]}(m\nu) \,,
\end{eqnarray} 
where $\alpha_s^{[3]}(m\nu)$ is the QCD coupling with 3-loop running.

\section{Anomalous dimensions for Production Currents} 
\label{sectioncurrent}

\begin{figure}
  \epsfxsize=14cm \centerline{\epsfbox{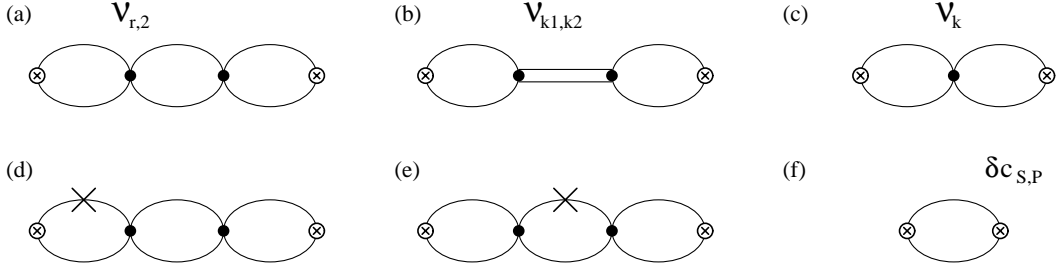}}
%  \vskip 3cm
  \medskip
{\caption{
Three-loop current correlator diagrams and the counterterm diagram for the
computation of the NLL anomalous dimensions of $c_{S,P}$.
}
\label{figcurrents}}
\end{figure}

\begin{figure}[!t]
\epsfxsize=10.cm 
\epsffile[80 430 560 720]{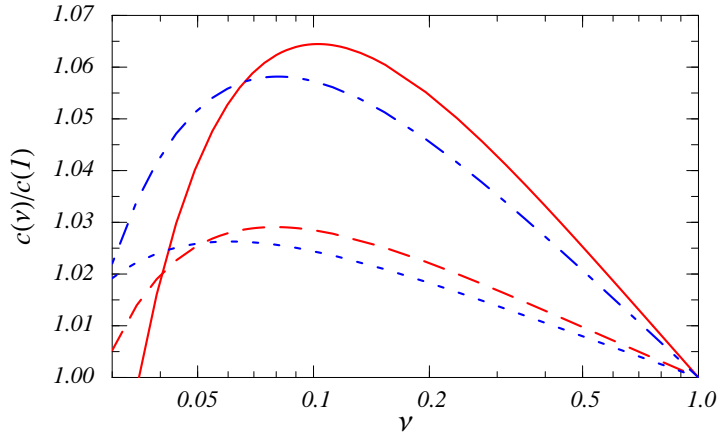}
%  \epsfxsize=12cm \centerline{\epsfbox{figures/cspplot.ps}}
%  \vskip 3cm
%  \medskip
{\caption{NLL running of the normalized Wilson coefficient for S-wave
    and P-wave currents for $m=220$~GeV (solid and dashed lines) and
    for $m=500$~GeV (dot-dashed and dotted lines).}
\label{cscpplot}}
\end{figure}

We consider the currents describing squark-antisquark production in S- and
P-wave states relevant for $\gamma\gamma$ and $e^+e^-$ collisions,
respectively:
\begin{eqnarray} 
\label{JSJP}
  J_{S,\bf p} = 
    \psi_{\bmp}^*\,\chi_{-\bmp}^*
   \,,\qquad
  {\bf J}_{P,\bf p}=   \psi_{\bmp}^*\,\bmp\,\chi_{-\bmp}^*\,,
\end{eqnarray}
where $c_S(\nu)$ and $c_P(\nu)$ are the corresponding Wilson coefficients.
Concerning the summation of large QCD logarithms the LL matching conditions
are conventionally normalized to unity. For $e^+e^-$ annihilation the NLL
order matching condition for the P-wave vector current can be obtained from 
matching to the one-loop QCD amplitudes. From the QCD one-loop results
obtained in~\cite{Drees,Beenakker} we find
\begin{eqnarray}
c_P(1) = 1-\frac{C_F \alpha_s(m)}{\pi}
\,.
\label{Pcurrentmatch}
\end{eqnarray} 
The corresponding matching condition for the S-wave current relevant
for $\gamma\gamma$ collisions is presently unknown.
In analogy to
the quark case the currents in Eqs.~(\ref{JSJP}) do not run at LL
order since there are in general no UV-divergences at one-loop in the
effective theory that can renormalize currents that are leading in $v$
for a particular quantum number. To determine the NLL anomalous
dimensions we compute the current correlators diagrams in
Fig.~\ref{figcurrents} rather than two-loop vertex diagrams.
The former method is more convenient since IR-divergences originating from the
Coulomb phase cancel automatically and a distinction between IR- and
UV-divergences becomes unnecessary~\cite{Hoang3loop}. In
Fig.~\ref{figcurrents} four-quark 
interactions without labels refer to the Coulomb potential and crossed circles
to the currents (and their complex conjugate). Crossed squark lines refer to
insertions of the NNLL order squark kinetic energy operator. Combinatorial
factors and diagrams obtained by flipping the graphs left-to-right and
up-to-down are understood. The results for the NLL order anomalous dimensions
read
\begin{eqnarray} 
\nu \frac{\partial}{\partial \nu} \ln[c_S(\nu)] 
& = &
-\:\frac{{\cal V}_c^{(s)}(\nu)
  }{ 16\pi^2} \bigg[ \frac{ {\cal V}_c^{(s)}(\nu)}{4}
  +{\cal V}_r^{(s)}(\nu) +{\cal V}_2^{(s)}(\nu) \bigg] 
\nn\\[2mm] &&
  +\, \frac{{\cal V}_k^{(s)}(\nu)}{2} 
 + \alpha_s^2(m\nu)
  \,[ 3 {\cal V}_{k1}^{(s)}(\nu) + 2 {\cal V}_{k2}^{(s)}(\nu) ] \,.
\nn\\[3mm]  
\frac{\partial}{\partial \nu} \ln[c_P(\nu)]
& = &  
-\:\frac{{\cal V}_c^{(s)}(\nu)
  }{ 48\pi^2} \bigg[ \frac{ {\cal V}_c^{(s)}(\nu) }{4}
 + {\cal V}_r^{(s)}(\nu) \bigg] 
\nn\\[3pt] && 
  +\, \frac{ {\cal V}_k^{(s)}(\nu) }{6} 
 + \alpha_s^2(m\nu)
  \,\bigg[ {\cal V}_{k1}^{(s)}(\nu) + \frac{2}{3} {\cal V}_{k2}^{(s)}(\nu)
    \bigg] \,.
\label{runc}
\end{eqnarray}
The expression for the S-wave current agrees with the one for the S-wave
current in the quark production case~\cite{LMR} (see also 
Refs.~\cite{HoangStewartultra,Pineda1}),
up to the absence of the spin-dependent
terms and the fact that the coefficient ${\cal V}_2^{(T)}$ differs for the
both cases. As the solution for Eqs.~(\ref{runc}) we find
\begin{eqnarray} 
\label{NLLc}
\ln \frac{c_S(\nu)}{c_S(1)}
  & = & b_2\,\pi \alpha_s(m)\, (1-z) + b_3\,\pi \alpha_s(m)\, \ln (z) 
 \nn\\[3pt] &&   + \,b_4\,\pi \alpha_s(m)\,\Big[1-z^{1-13C_A/(6\beta_0)} \Big] 
  + b_0\,\pi \alpha_s(m)\Big[z-1-w^{-1}\ln (w)\Big]\,, 
\nn\\[3mm]
\ln \frac{c_P(\nu)}{c_P(1)}
  & = & d_2\,\pi \alpha_s(m)\, (1-z) +
  d_0\,\pi \alpha_s(m)\Big[z-1-w^{-1}\ln (w)\Big]\,, 
%  & = & \frac{\pi\, C_F (C_A+2C_F)}{3\beta_0} \, (1-z)\, \alpha_s(m) \nn\\[3pt]
%  && + \frac{8\pi\, C_A C_F}{9\, \beta_0}
%  (C_A+4C_F)\,\alpha_s(m)\Big[z-1-w^{-1}\ln (w)\Big]\,, 
\end{eqnarray}
where
\begin{eqnarray}
b_2 & = & \frac{C_F \,[\, C_A C_F(9C_A-100C_F)-\beta_0(26C_A^2+19C_F
    C_A -32C_F^2)\,]\,}{26\beta_0^2 C_A} 
\,,\nn\\[3mm]
b_3 & = & 
-\frac{C_F^2\,[\,C_A(9C_A - 100 C_F) - 6 \beta_0(3C_A - 4C_F)}{2 \beta_0^2(6\beta_0 - 13 C_A)}
%\frac{C_F^2 \,[\, 9C_A(2\beta_0-C_A)+4C_F(25C_A-6\beta_0) \,]\,}{2\beta_0^2(6\beta_0-13C_A)}
\,,\nn\\[3mm] 
b_4 & = & \frac{24C_F^2(11C_A-3\beta_0)(5C_A+8C_F)}{13C_A(6\beta_0-13C_A)^2} 
\,,\nn\\[3mm] 
b_0 & = & - \frac{8C_F(C_A+C_F)(C_A+2C_F)}{3\beta_0^2}
\,,\nn\\[3mm]
d_2 & = & -\frac{ C_F (C_A+2C_F)}{3\beta_0}
\,,\nn\\[3mm]
d_0 & = & -\frac{8 C_A C_F(C_A+4C_F)}{9\beta_0^2}
\end{eqnarray}
In Fig.~{\ref{cscpplot}} we have displayed the NLL running of the 
Wilson coefficients $c_S(\nu)$ and $c_P(\nu)$ normalized to their
matching values. For the input parameters we have chosen  
two different values for the heavy scalar mass, $m=220$~GeV and
$m=500$~GeV, and $\alpha_s(m_Z)=0.118$, taking leading-logarithmic
running for $\alpha_s$ with $n_f=5$ active massless quark flavors and
no active massless squarks ($n_s=0$). We find that the $\nu$-variation
for the S-wave coefficient is much stronger than for the P-wave
coefficient. While the S-wave coefficient increases by about 6\% for 
$\nu\sim\alpha_s$ compared to the matching value at $\nu=1$, the
P-wave coefficient only increases by less than 3\%.
Furthermore, for S- and P-wave coefficients the maximum slightly
decreases with the heavy scalar mass and also moves towards smaller
values of $\nu$. The latter feature can be understood qualitatively
from the fact that the average velocity $\langle
v\rangle\sim\alpha_s(m\alpha_s)$ decreases with the heavy scalar mass
$m$. 

\section{Conclusion}
\label{sectionconclusion}
In this work we have presented the vNRQCD Lagrangian for a particle-antiparticle
pair of colored scalars in
the fundamental representation of QCD. The paper provides the needed ingredients for a NLL order
description of the QCD effects of scalar-antiscalar production close to threshold
in $e^+e^-$ and $\gamma\gamma$ collisions,
and for scalar-antiscalar bound state quarks
energies at NNLL order.

The proposed Lagrangian is the spinless counterpart of the vNRQCD Lagrangian 
first formulated for quarks. 
To the order required in this work, the ultrasoft
sector of the quark and scalar theories agrees, since to the order we are
interested interactions with ultrasoft 
energy gluons are not sensitive to the spin. As a consequence the 1-loop ultrasoft
renormalization is identical in both theories.
We also find that the soft coefficients in the scalar theory are in complete 
analogy with the corresponding ones in the quark theory. This is related to
similarities in the structure of the HQET and HSET Lagrangians up to order $1/m$
due to reparametrization invariance.
Differences arise in matching conditions that are not protected by symmetries
and in the running induced by soft divergences
since for scalars there are no mixing due to spin-dependent operators.

\begin{acknowledgments} 
We thank A.~Manohar for comments to the manuscript, and T.~Teubner for
discussions. 

\end{acknowledgments}

%\newpage
\appendix
\section{Construction of the tree level HSET Lagrangian}
\label{HSETconstruction}

The QCD Lagrangian for a colored scalar field $\phi$ with mass $m$ reads 
\begin{eqnarray}
{\cal L}_{\rm{sQCD}}=\big(D_{\mu}\phi\big)^*D^{\mu}\phi-m^2\phi^{*}\phi \,.
\end{eqnarray}
To obtain the tree level heavy scalar effective Lagrangian from ${\cal
  L}_{\rm{QCD}}$ we need to take the limit $m\to \infty$ holding the
  heavy scalar's four-velocity $v^\mu$ fixed ($v^2=1$). 
With this aim, the colored scalar field with momentum $p=mv+k$, where $k$ is small,
can be decomposed into two pieces, 
\begin{eqnarray}
\phi(x)=\frac{e^{-im\,v\cdot x}}{\sqrt{2m}}\left( \psi+\chi \right) \,,
\label{decom}
\end{eqnarray}
\begin{eqnarray}
\psi(x)=e^{im\,v\cdot x}P_{+}\,\phi(x) \quad,\quad \chi(x)=e^{im\,v\cdot x}P_{-}
\,\phi(x) \,,
\label{components}
\end{eqnarray}
where 
\begin{eqnarray}
P_{\pm}={1 \over \sqrt{2m}} \left(\pm i\,v\cdot D+m \right) \,,
\label{proj}
\end{eqnarray}
project out the positive (particle) and negative (antiparticle) energy
components of the field in the infinite mass 
limit. 
The explicit phase-factor introduced in Eq.~(\ref{components})
extracts the relativistic fluctuations of the particle field.  
This procedure is in complete analogy with the decomposition performed
for a heavy quark field  to write down the HQET
Lagrangian, where the upper and lower components of the
heavy quark field are projected out by $P_{\pm}=(1\pm \slash{v})/2$. 
The $\sqrt{2m}$ factor has been introduced 
to achieve the non-relativistic normalization for the fields $\psi$ and $\chi$.  

The QCD Lagrangian with the decomposition above reads 
\begin{eqnarray}
{\cal L}_{\rm{sQCD}} & = &
(\psi^*+\chi^*)\,\left[ \,iv\cdot D +{(iv\cdot D)^2 \over 2m} 
-{ D^2_\perp\over 2m}\, \right]\,(\psi+\chi)
\nn\\[3pt] 
 & = & \psi^*\Big( iv\cdot D-{D_\perp^2 \over 2m} \Big)\psi
 - \chi^*\Big( iv\cdot D+2m+{D_\perp^2\over 2m} \Big)\chi 
 - \chi^*\,{D_\perp^2 \over 2m}\, \psi 
 - \psi^*\,{D_\perp^2 \over 2m}\, \chi \,,
 \label{QCDLag}
\end{eqnarray}
where $D_\perp^{\mu}\equiv D^\mu -v^{\mu}(v\cdot D)$ is the component
of the covariant derivative orthogonal to the velocity, i.e. $v\cdot
D_\perp=0$. In the second line of Eq.~(\ref{QCDLag}) we have
eliminated the term $(i v\cdot D)^2$ and the temporal derivatives in
the crossed terms with the help of
Eqs.~(\ref{decom}-\ref{proj}). (Doing the manipulations for $\psi$ and
$\chi$ separately also the equation of motion for the field $\phi$ in
the full theory is required.) The field $\psi$  describes a massless
degree of freedom, while $\chi$ corresponds to fluctuations with twice
the heavy scalar mass and can be integrated out of the theory at the
energy scales we are interested in. At tree-level this can be done 
by solving the equation of motion for the field $\chi$:
\begin{eqnarray}
\Big(2m+iv\cdot D+{D_\perp^2 \over 2m}\Big) \chi & = & 
-{D_\perp^2 \over 2m}\,\psi\,,
\label{eom}
\end{eqnarray}
which implies
\begin{eqnarray}
 \chi & = & -{ 1\over 2m+iv\cdot D+{D_\perp^2 \over 2m}}\,
{D_\perp^2 \over 2m}\,\psi \,\simeq\, 
\bigg( {iv\cdot D \over 2m}+{D_\perp^2 \over (2m)^2}
+\dots \bigg){D_\perp^2 \over (2m)^2}\psi\,.
\label{eomexp}
\end{eqnarray}
The expansion above is justified because derivatives acting on $\psi$
are of the order of the residual momentum of the heavy scalar,
$k=p-mv\ll m$. Inserting the relation~(\ref{QCDLag}) in the QCD
Lagrangian gives a local effective Lagrangian for heavy scalars as an
expansion in $1/m$: 
\begin{eqnarray}
{\cal L}_{\rm{HSET}} & = &
\psi^*\left( \,iv\cdot D -{D_\perp^2\over 2m}
+{ D^4_\perp\over 8m^3}-{ D^2_\perp\over 4m^2}\,iv\cdot D\,{ D^2_\perp\over 4m^2}
+{\cal O}\Big(\frac{1}{m^5}\Big)
\, \right)\psi
\,.
 \label{HSET}
\end{eqnarray}
The field redefinition
\begin{eqnarray}
\psi\,\to\,\bigg(1+{D^4_\perp\over 32m^4}\bigg)\psi
\label{field_red}
\end{eqnarray}
can be used to eliminate the ${\cal O}(1/m^4)$ time derivative term acting on $\psi$ and
cast the tree level HSET Lagrangian in the form
\begin{eqnarray}
{\cal L}_{\rm{HSET}} & = &
\psi^*\left( \,iv\cdot D -{D_\perp^2\over 2m}
+{ D^4_\perp\over 8m^3}
+\Big\{\big[iv\cdot D,{D_\perp^2\over 8m^2}\big],{D_\perp^2\over 4m^2}\Big\}
+{\cal O}\Big(\frac{1}{m^5}\Big)\, \right)\psi
\nonumber\\[3mm] & = &
\psi^*\left( \,iv\cdot D -{D_\perp^2\over 2m}
+{ D^4_\perp\over 8m^3}
-g \frac{v_\alpha \{\{G^{\alpha\mu},D^\perp_\mu\},D_\perp^2\}
}{32 m^4}
+{\cal O}\Big(\frac{1}{m^5}\Big)\, \right)\psi
\,.
 \label{HSET2}
\end{eqnarray}
In the heavy scalar rest frame, $v=(1,0,0,0)$, $D_\perp^{\mu}=(0,-\bmD)$,
the tree level HSET Lagrangian takes the form
\begin{eqnarray}
{\cal L}_{\rm{HSET}} & = &
\psi^*\left( \,iD_0 +{{\bmD}^2\over 2m}
+{ {\bmD}^4\over 8m^3}
-g \,\frac{\{\{\bmD,\bmE\},\bmD^2\}  }{32 m^4}
+{\cal O}\Big(\frac{1}{m^5}\Big)\, \right)\psi
\,.
 \label{HSET-NRQCD}
\end{eqnarray}
Note that at tree level the Darwin term $[ {\bmD \cdot \bmE} ]/(8
m^2)$ is absent, and at order $1/m^3$ only the kinetic 
${\bmD}^4/8m^3$ operator exists. The full operator basis up to order
$1/m^3$ (relevant for HSET and HQET) can be found in
Ref.~\cite{Manoharm3}.

%%%%%%%%%%%%%%%%%%%%%%%%%%%%%%%%%%%%%%%%%%%%%%%%%%%%%%%%%%

%Bibliography

\end{document}